\documentclass[%
prl%
 ,twocolumn%
 ,superscriptaddress%
,tightenlines%
,amssymb,amsmath,nobibnotes, aps, prl]{revtex4}
\usepackage{dcolumn}
\usepackage{bm}
\usepackage[dvips]{graphicx}
\begin{document}
\title{Electric-Field-Induced Nuclear Spin Resonance Mediated by Oscillating Electron Spin Domains in GaAs-Based Semiconductors}
\author{N. Kumada}
\affiliation{NTT\,Basic\,Research\,Laboratories,\,NTT\,Corporation,\,3-1\,Morinosato-Wakamiya,\,Atsugi\,243-0198,\,Japan}
\author{T. Kamada}
\affiliation{NTT\,Basic\,Research\,Laboratories,\,NTT\,Corporation,\,3-1\,Morinosato-Wakamiya,\,Atsugi\,243-0198,\,Japan}
\affiliation{Department of Physics, Tohoku University, Sendai 980-8578, Japan}
\author{S. Miyashita}
\affiliation{NTT-AT, 3-1 Morinosato-Wakamiya, Atsugi 243-0198, Japan}
\author{Y. Hirayama}
\affiliation{Department of Physics, Tohoku University, Sendai 980-8578, Japan}
\affiliation{ERATO Nuclear Spin Electronics Project, Sendai 980-8578, Japan}
\author{T. Fujisawa}
\affiliation{NTT\,Basic\,Research\,Laboratories,\,NTT\,Corporation,\,3-1\,Morinosato-Wakamiya,\,Atsugi\,243-0198,\,Japan}

\date{Version: \today}

\begin{abstract}
We demonstrate an alternative nuclear spin resonance using radio frequency (RF) electric field (nuclear electric resonance: NER) instead of magnetic field.
The NER is based on the electronic control of electron spins forming a domain structure.
The RF electric field applied to a gate excites spatial oscillations of the domain walls and thus temporal oscillations of the hyperfine field to nuclear spins.
The RF power and burst duration dependence of the NER spectrum provides insight into the interplay between nuclear spins and the oscillating domain walls.

\end{abstract}
\maketitle


In GaAs-based semiconductors, electronic and nuclear spins are coupled with each other through the contact hyperfine interaction.
This allows nuclear spins to be dynamically polarized by manipulating electron spins using electrical \cite{Wald,Kronmuller1999,OnoPRL,Kawamura} or optical \cite{Bracker} means.
In two-dimensional systems, a combination of the dynamic nuclear spin polarization and the irradiation of radio frequency (RF) magnetic field is used to control nuclear spins coherently in submicron regions \cite{Machida,YusaNature,SanadaPRL2} or to perform nuclear magnetic resonance (NMR) for probing electronic properties in quantum Hall (QH) systems \cite{OStern,KumadaPRL3}.
In these experiments, nuclear spin resonance is induced by RF magnetic field generated by a coil wound around the sample or a strip line fabricated on the sample.
An alternative is the electric-field-induced nuclear spin resonance (nuclear electric resonance: NER).
Since oscillating electric field can be generated by exciting a gate, it has an advantage of spatial selectivity.
The advantage of NER combined with a scanning gate technique \cite{Topinka} would allow local probing of electronic properties or nuclear resonance imaging with nanoscale resolution.

For electronic systems, single-spin resonance utilizing the spatial selectivity of an oscillating electric field has been proposed \cite{Debald,Tokura,Golovach,Flindt} and demonstrated \cite{Nowack,Laird}.
In these studies, electron motion induced by an oscillating electric field is coupled to the spin through the spin-orbit interaction \cite{Debald,Golovach,Flindt,Nowack}, a slanting magnetic field \cite{Tokura} or an inhomogeneous hyperfine field \cite{Laird}.
For nuclear systems, however, the techniques explotting spin-motion coupling can not be applied because nuclei are fixed at their lattice sites.
So far, there is only a proposal \cite{Aronov} of NER mediated by the electron-spin-orbit and hyperfine interactions.




In this Letter, we propose and demonstrate NER exploitting electronic control of the hyperfine field from electron spin domains.
We use current-induced nuclear spin polarization along domain walls in a QH system formed in a GaAs quantum well \cite{Kronmuller1999}.
After the nuclear spin polarization, a burst of RF electric field is applied to a gate covering the whole area of the device.
When the RF frequency is set at the nuclear resonance frequency, nuclear spins are depolarized.
The RF power and burst duration dependence of the NER spectrum indicates that the NER is caused by temporal oscillations of the hyperfine field due to spatial oscillations of the domain walls excited by the RF electric field.

\begin{figure}[t]
\begin{center}
\includegraphics[width=1.0\linewidth]{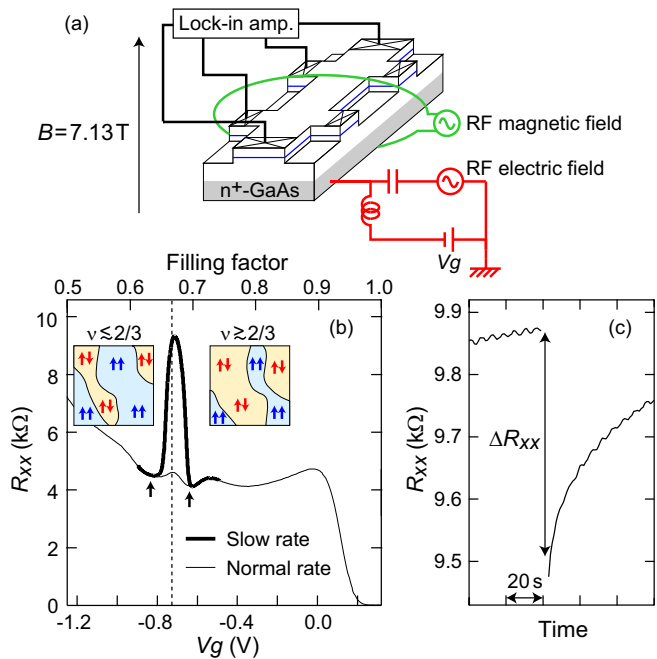}
\caption{
(Color online).
(a) Schematic of the experimental setup.
(b) $R_{xx}$ at $B=7.13$\,T obtained by sweeping the gate bias at normal ($dV_g/dt=5.0\times 10^{-3}$\,V/s) and slow ($dV_g/dt=5.0\times 10^{-5}$\,V/s) sweep rates.
The filling factor is shown on the top axis.
The dotted line represents $\nu =2/3$.
Arrows point to $R_{xx}$ minima for the spin-polarized and -unpolarized states.
Insets show illustrations of domain structures for filling factors slightly higher and lower than $\nu =2/3$, where the spin-unpolarized and -polarized states are dominant, respectively.
(c) Time evolution of $R_{xx}$ at $\nu =2/3$ ($V_g=-0.74$\,V).
The resistance drop is caused by the burst of the RF electric field at 51.971\,MHz.
The burst duration and the output power of the RF generator are $\tau _{\rm RF}=1$\,s and $P_{\rm RF}=-15$\,dBm, respectively.
}
\label{demo}
\end{center}
\end{figure}

The experimental setup is schematically shown in Fig.\,\ref{demo}(a).
The sample, which consists of a 20-nm-wide GaAs quantum well, was processed into a 50-$\mu $m-wide Hall bar.
The low-temperature mobility is 2.3$\times 10^{6}$\,cm$^2/$V\,s at an electron density of $1.7\times 10^{11}$\,cm$^{-2}$.
The density can be controlled using an $n^+$-GaAs substrate acting as a gate.
The sample was mounted in a dilution refrigerator with a base temperature of 100\,mK.
The longitudinal resistance $R_{xx}$ was measured by a standard low-frequency lock-in technique.
A relatively large current of $100$\,nA was used for the dynamic nuclear spin polarization \cite{Kraus}.
Throughout this work, we fixed the static magnetic field $B=7.13$\,T perpendicular to the two-dimensional electron system.
RF electric field for the NER is applied to the gate through a bias-tee.
To perform NMR as a reference, a wire loop for generating RF magnetic field was placed around the sample.

We use current-induced and resistively detected nuclear spin polarization \cite{Kronmuller1999}, which occurs at the phase transition point between the spin-polarized and -unpolarized states of the QH system at Landau level filling factor $\nu =2/3$; at the transition point, a domain structure of the two phases is formed \cite{Kraus,Verdene} and flip-flop scattering upon transport across the domain walls produces nuclear spin polarization, which acts as an additional source of scattering and can be detected via $R_{xx}$ \cite{Kraus,HashimotoPRB}.
Figure\,\ref{demo}(b) presents $R_{xx}$ obtained by sweeping the gate bias $V_g$ at normal and slow sweep rates.
At the two $R_{xx}$ minima at $V_g=-0.84$\,V and $-0.64$\,V indicated by arrows, the spin-polarized and -unpolarized states are formed, respectively.
Nuclear spin polarization is manifested as the slow enhancement of $R_{xx}$ between the two minima, where the domain structure is formed.
In this system, the domain structure depends on $V_g$ as illustrated in the inset of Fig.\,\ref{demo}(b).
Therefore, the application of an RF electric field to the gate excites spatial oscillations of the domain walls and thus temporal oscillations of the hyperfine field to nuclear spins at the RF frequency.
Since the hyperfine field in the domain walls has an in-plane component perpendicular to the static magnetic field, its oscillation at the resonance frequency can induce NER.

Figure\,\ref{demo}(c) demonstrates the response of nuclear spins to the RF electric field applied to the gate.
Prior to the RF application, nuclear spins are polarized at $\nu =2/3$ ($V_g=-0.74$\,V), where $R_{xx}$ increases slowly (not shown here).
Then the RF burst at a frequency $f_{\rm RF}$ is applied to the gate, which causes the depolarization of nuclear spins as manifested by the $R_{xx}$ drop.
Here, the size of the resistance drop $\Delta R_{xx}$ is defined by subtracting the average of $R_{xx}$ over 5\,s just after the RF application from that before.

\begin{figure}[t]
\begin{center}
\includegraphics[width=0.7\linewidth]{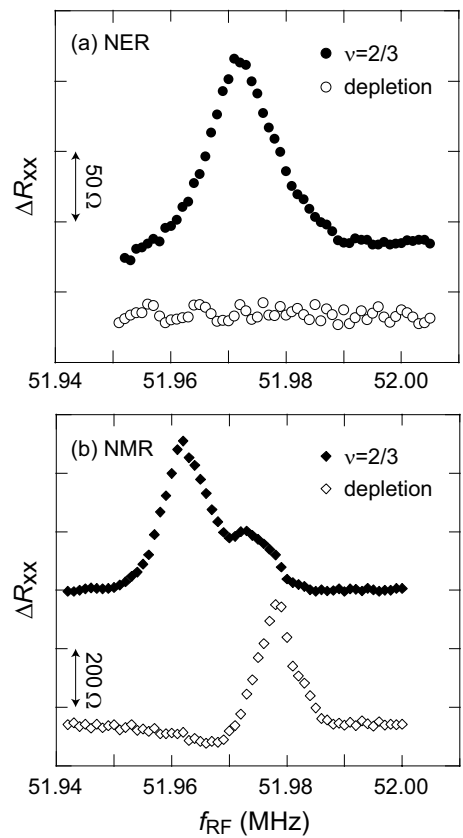}
\caption{
(a) NER spectrum of $^{75}$As for the $\nu =2/3$ system (solid circles).
$\Delta R_{xx}$ caused by the RF electric field of $\tau _{\rm RF}=1$\,s and $P_{\rm RF}=-15$\,dBm is plotted as a function of $f_{\rm RF}$.
$\Delta R_{xx}$ at depletion (open circles) is also plotted.
(b) Resistively-detected NMR spectra of $^{75}$As for the $\nu =2/3$ system (solid diamonds) and at depletion (open diamonds) obtained by the irradiation of the RF magnetic field of $\tau _{\rm RF}=300$\,ms and $P_{\rm RF}=10$\,dBm.
Traces are vertically offset for clarity.
}
\label{spectra}
\end{center}
\end{figure}

In Fig.\,\ref{spectra}(a), $\Delta R_{xx}$ is plotted as a function of $f_{\rm RF}$ (solid circles).
The spectrum shows a peak at $f_{\rm RF}=51.971$\,MHz, which corresponds to the resonance frequency of $^{75}$As at $B=7.13$\,T.
To rule out unintentional irradiation of the RF magnetic field generated by the current in the leads, we carried out a similar measurement while depleting electrons by a large negative static bias $V_g=-2.3$\,V during the RF application.
$\Delta R_{xx}$ at depletion [open circles in Fig.\,\ref{spectra}(a)] does not show any resonance feature, indicating that the observed nuclear spin resonance for the $\nu =2/3$ system is indeed electric field induced.

\begin{figure}[t]
\begin{center}
\includegraphics[width=0.76\linewidth]{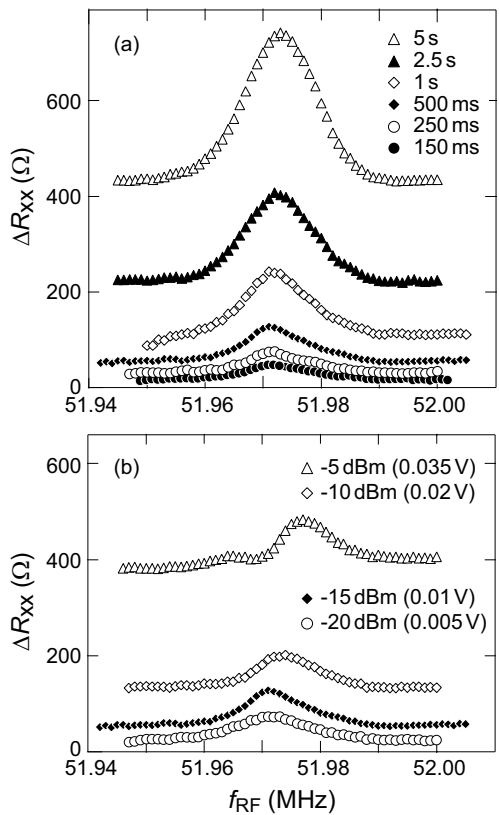}
\caption{
NER spectra of $^{75}$As for the $\nu =2/3$ system for several values of (a) $\tau _{\rm RF}$ at $P_{\rm RF}=-15$\,dBm and (b) $P_{\rm RF}$ at $\tau _{\rm RF}=500$\,msec, respectively.
The amplitude of the bias modulation for each $P_{\rm RF}$ is shown inside the parentheses.
}
\label{durationpower}
\end{center}
\end{figure}

The NER spectrum for the $\nu =2/3$ system is compared to the resistively detected NMR spectra, which are obtained by applying RF magnetic field \cite{KumadaPRL3}.
Figure\,\ref{spectra}(b) shows the NMR spectra for the $\nu =2/3$ system (solid diamonds) and at depletion (open diamonds).
The NMR spectrum at depletion gives a resonance frequency of nuclei free from the hyperfine interaction.
In the NMR spectrum for the $\nu =2/3$ system, two peaks appear because of the spatial variation of the hyperfine field parallel to the static magnetic field in the domain structure: the right peak, which appears at a similar frequency to the NMR frequency at depletion, comes from nuclei interacting with spin-unpolarized domains, whereas the left peak with a Knight shift of $\sim 17$\,kHz from the spin-polarized domains \cite{OStern,smalldifference}.
The frequency of the single NER peak lies between the two NMR peaks.
The difference in the resonance frequency indicates that for the NER the picture that the nuclear spin resonance occurs in the spin-polarized and -unpolarized domains is no longer valid.

\begin{figure}[t]
\begin{center}
\includegraphics[width=0.88\linewidth]{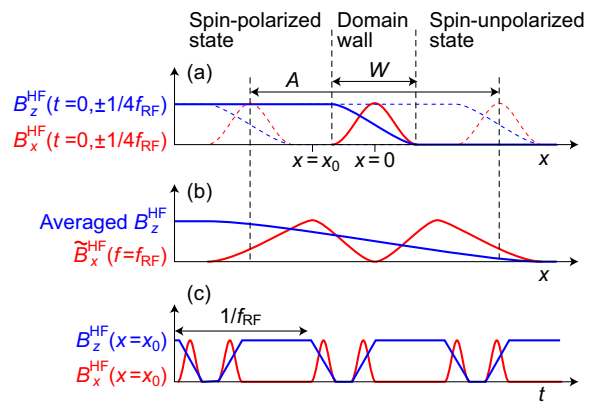}
\caption{(Color online).
(a) $x$ and $z$ components of the hyperfine field, $B^{\rm HF}_x$ and $B^{\rm HF}_z$ respectively, as a function of the distance from a domain wall $x$ at three moments in time $t=0, \pm 1/4f_{\rm RF}$.
(b) Time-averaged $B^{\rm HF}_z$ and the spectral density of $\tilde{B}^{\rm HF}_x(f=f_{\rm RF})$ at $f_{\rm RF}$.
(c) $B^{\rm HF}_x$ and $B^{\rm HF}_z$ at $x=x_0$ as a function of time $t$.
}
\label{hyperfinefield}
\end{center}
\end{figure}

The manner in which the NER spectrum depends on the burst duration $\tau _{\rm RF}$ and output power of the RF generator $P_{\rm RF}$ reveals the mechanism of the NER.
Here, the amplitude of the bias modulation for $P_{\rm RF}=-20$\,dBm, $-15$\,dBm, $-10$\,dBm and $-5$\,dBm is roughly estimated to be 0.005\,V, 0.01\,V, 0.02\,V and 0.035\,V, respectively \cite{amplitude}.
Figures\,\ref{durationpower}(a) and (b) show the NER spectra for several values of $\tau _{\rm RF}$ and $P_{\rm RF}$, respectively.
The vertical shifts of the spectra are due to heating, which increases with the integrated RF power $\tau _{\rm RF}\times P_{\rm RF}$.
As $\tau _{\rm RF}$ is increased, the resonance strength increases monotonically with a long time constant of the order of 1\,s.
As $P_{\rm RF}$ is increased, on the other hand, the resonance strength slightly increases and then saturates at a smaller value for $P_{\rm RF}\geq -15$\,dBm.
In addition, for larger $P_{\rm RF}$, the peak splits into two peaks.

The $P_{\rm RF}$ and $\tau _{\rm RF}$ dependence can be explained by the spatial and time dependence of the hyperfine field.
Consider a domain wall with a width $W$ centered at $x=0$ between the spin-polarized ($x<0$) and -unpolarized ($x>0$) regions.
Corresponding hyperfine field, $B^{\rm HF}_x$ and $B^{\rm HF}_z$ respectively for the in-plane ($x$) and out-of-plane ($z$) components, is illustrated as a function of $x$ in Fig.\,\ref{hyperfinefield}(a).
$B^{\rm HF}_x$ is finite only in the domain wall, while $B^{\rm HF}_z$ changes between the values for the full and null spin polarizations across the domain wall.
Note that $W$ is estimated to be $\sim 100$\,nm \cite{Shibata}.
Under the RF electric field, the domain wall oscillates spatially with an amplitude $A$ determined by $P_{\rm RF}$ [dotted vertical lines in Fig\,\ref{hyperfinefield}(a)], which causes temporal oscillations of $B^{\rm HF}_x(t)$ and $B^{\rm HF}_z(t)$ at a position near the domain wall [Fig\,\ref{hyperfinefield}(c)].

The resonance strength is determined by the spectral density of $B^{\rm HF}_x(t)$ at the resonance frequency.
The amplitude of the oscillation of $B^{\rm HF}_x(t)$ increases with $P_{\rm RF}$ until $A$ reaches $W$, above which it is limited by the strength of the hyperfine interaction independent of $P_{\rm RF}$ [Fig\,\ref{hyperfinefield}(c)].
As a result, the resonance strength saturates for larger $P_{\rm RF}$.
This implies that if the relation between $P_{\rm RF}$ and $A$ is known, $W$ can be obtained from the $P_{\rm RF}$ dependence of the resonance strength.
The slow $\tau _{\rm RF}$ dependence is due to the small Fourier component of $\tilde{B}^{\rm HF}_x(f=f_{\rm RF})$ at $f_{\rm RF}$: the Fourier spectrum $\tilde{B}^{\rm HF}_x(f)$ contains not only a fundamental at $f_{\rm RF}$ but also large amount of harmonics because $B^{\rm HF}_x(t)$ is not a sinusoidal function  [Fig\,\ref{hyperfinefield}(c)].

The resonance frequency reflects the Knight shift, which corresponds to the time-averaged $B^{\rm HF}_z$.
When $P_{\rm RF}$ is large so that $A>W$, $B^{\rm HF}_z(t)$ follows a square function with a duty ratio depending on $x$ [Fig.\,\ref{hyperfinefield}(c)], where the Knight shift changes almost linearly between full and zero for $|x|<A/2$ [Fig.\,\ref{hyperfinefield}(b)].
In such a situation, the spatial distribution of $\tilde{B}^{\rm HF}_x(x,f=f_{\rm RF})$ modifies the shape of the spectrum through the Knight shift gradient.
$B^{\rm HF}_x(t)$ follows peaks whose interval depends on $x$ with the repetition rate $f_{\rm RF}$ [Fig.\,\ref{hyperfinefield}(c)].
$\tilde{B}^{\rm HF}_x(f=f_{\rm RF})$ is finite for $|x|<A/2$ except at $x=0$ [Fig.\,\ref{hyperfinefield}(b)]: at $x=0$, the interval between the peaks is uniform at $1/2f_{\rm RF}$, where the fundamental is $2f_{\rm RF}$ and $\tilde{B}^{\rm HF}_x(f=f_{\rm RF})$ is zero.
As a result, $\tilde{B}^{\rm HF}_x(x,f=f_{\rm RF})$ has two peaks, leading to the two resonance peaks.
Note that the nuclear spin polarization is expected to be inhomogeneous across the domain wall and the distribution also modifies the spectrum.
For $A<W$, $B^{\rm HF}_z(t)$ is determined by the electron spin configuration in the domain wall and does not follow the square function.
The single resonance peak for smaller $P_{\rm RF}$ suggests that $B^{\rm HF}_z(t)$ fluctuates by electron spin fluctuations in the domain wall and the Knight shift is averaged out.


\begin{figure}[t]
\begin{center}
\includegraphics[width=0.9\linewidth]{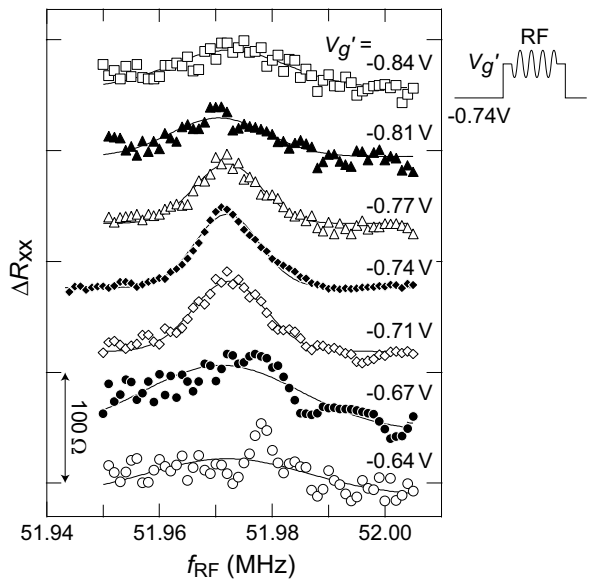}
\caption{
NER spectra of $^{75}$As for several values of $V_g'$ for $\tau _{\rm RF}=500$\,ms and $P_{\rm RF}=-15$\,dBm.
Experimental sequence is shown on the right.
Lines are guides to the eye.
Traces are vertically offset for clarity.
}
\label{electronicsystem}
\end{center}
\end{figure}


Finally, we show NER spectra obtained by shifting the position of the domain walls during the RF application.
The NER spectra in Fig.\,\ref{electronicsystem} are obtained by setting the gate bias to a temporal value $V_g'$ during the RF application, while the nuclear spin polarization and its resistive detection are still carried out at $\nu =2/3$ ($V_g=-0.74$\,V).
The shift of domain walls increases with $|V_g'+0.74\,{\rm V}|$ \cite{percolation}.
When the shift is small ($|V_g'+0.74\,{\rm V}|\leq 0.03$\,V), the NER spectrum shows a clear resonance.
For larger $|V_g'+0.74\,{\rm V}|\geq 0.07$\,V, on the other hand, the resonance peak almost disappears.
These results indicate that the NER is sensitive to the nuclear spin polarization along the domain walls.
We suggest that the NER combined with a scanning gate technique can detect the position of the domain walls and thereby allows imaging of the domain structure.


\begin{acknowledgments}
The authors are grateful to T. Ota, K. Takashina and K. Muraki for fruitful discussions and T. Kobayashi for sample processing.
This work was supported in part by Grant-in-Aid for Scientific Research from the Ministry of Education, Culture, Sports, Science and Technology (No. 19204033).
\end{acknowledgments}


\begin{thebibliography}{99}
\bibitem{smalldifference}
Small difference in the NMR frequency for the spin-unpolarized domains and at depletion would be due to thermal spin excitations in the spin-unpolarized domains.
\bibitem{amplitude}
The amplitude of the bias modulation $V_{\rm RF}$ is obtained by measuring $R_{xx}$ while applying the nonresonant RF electric field: $R_{xx}$ becomes an average of the values for the bias between $V_g\pm V_{\rm RF}$.
\bibitem{percolation}
The domain structure disappears for $|V_g'+0.74\,{\rm V}|\geq 0.1$\,V, where the current-induced nuclear spin polarization does not occur [Fig.\,\ref{demo}(b)].
\end{thebibliography}
\end{document}